\documentclass[%
 aip,
 amsmath,amssymb,
 reprint,%
]{revtex4-1}

\usepackage{graphicx}
\usepackage{dcolumn}
\usepackage{bm}

\usepackage[utf8]{inputenc}
\usepackage[T1]{fontenc}
\usepackage{mathptmx}
\usepackage{epstopdf}
\usepackage{xcolor}

\begin{document}

\preprint{AIP/123-QED}

\title{Towards Surface Diffusion Potential Mapping on Atomic Length Scale}

\author{R. Villarreal}
\affiliation{Department of Quantum Matter Physics, University of Geneva, 24 Quai Ernest-Ansermet, CH-1211 Geneva 4, Switzerland}
\author{Christopher J. Kirkham}
\affiliation{National Institute for Materials Science (NIMS), Namiki 1-1, Tsukuba, Ibaraki, 305-0044, Japan}
\affiliation{London Centre for Nanotechnology and Department of Physics and Astronomy,University College London, London WC1E 6BT, United Kingdom}
\author{Alessandro Scarfato}
\affiliation{Department of Quantum Matter Physics, University of Geneva, 24 Quai Ernest-Ansermet, CH-1211 Geneva 4, Switzerland}
\author{David R. Bowler}
\affiliation{International Center for Materials Nanoarchitectonics (MANA), National Institute for Materials Science (NIMS), Namiki 1-1, Tsukuba, Ibaraki, 305-0044, Japan}
\affiliation{London Centre for Nanotechnology and Department of Physics and Astronomy,University College London, London WC1E 6BT, United Kingdom}
\author{Christoph Renner}
\email{Christoph.Renner@unige.ch}
\affiliation{Department of Quantum Matter Physics, University of Geneva, 24 Quai Ernest-Ansermet, CH-1211 Geneva 4, Switzerland}


\begin{abstract}
The surface diffusion potential landscape plays an essential role in a number of physical and chemical processes such as self-assembly and catalysis. Diffusion energy barriers can be calculated theoretically for simple systems, but there is currently no experimental technique to systematically measure them on the relevant atomic length scale. Here, we introduce an atomic force microscopy based method to semi-quantitatively map the surface diffusion potential on an atomic length scale. In this proof of concept experiment, we show that the atomic force microscope damping signal at constant frequency-shift can be linked to non-conservative processes associated with the lowering of energy barriers and compared with calculated single-atom diffusion energy barriers.
\end{abstract}

\maketitle


\section{\label{sec:level1}Introduction}

The behaviour of adspecies deposited on a surface is crucial for physico-chemical surface processes such as catalysis~\cite{over00,somorjai06,chen16,jones16}, surface decoration~\cite{dvorak16} and self-assembly~\cite{barth05} to name a few. For example, the emerging field of single-atom catalysts is currently facing two grand challenges: the proper surface decoration with single-atom adsorbates~\cite{zhu17} and the anchoring stability (low diffusion) of the single-atom catalysts at the atomic level~\cite{liu16,chen17}. Therefore, a good understanding of the surface diffusion potential is desirable for the adequate tuning of such processes.

As adspecies diffuse on a surface, they ultimately relax at local minima known as the adsorption sites. For a given stabilized adspecies, the minimum energy differences between these sites, preventing any further diffusion, are known as the potential energy barriers (or activation energies). Such barriers can be overcome globally, by increasing the adspecies energy with temperature~\cite{roder93}, or locally by lowering the confining barrier with, for instance, a scanning probe tip (stimulated diffusion)~\cite{sugimoto05}. 

Here, we focus on scanning probe microscopy techniques which have been used to characterize the surface diffusion of atoms and molecules on metals,~\cite{ehrlich66,ehrlich80,gomer90,gravil96,barth00,ternes08,hedgeland16} insulators~\cite{gravil96,repp04,sonnleitner11,repp16} and semiconductors~\cite{suliga83,song16}. Two main approaches have been adopted for determining associated energy barriers. In the first method, the adspecies thermally diffuse and their positions are tracked in time sequenced topographic images~\cite{gomer90} or with an atom-tracking technique~\cite{swartzentruber96}. These images allow to calculate the hopping rates at different temperatures via an Arrhenius analysis. However, this method is very time consuming, and the diffusion of specific individual adsorbates can be hard to track. The second approach is known as the onset method, where the energy barriers are obtained from an assumed growth model and experimentally observed island coverage~\cite{bott96,loske10}. More recently, several alternative methods have been proposed: non thermal diffusion stimulated by a tip probe (e.g. by inelastic excitations)~\cite{sonnleitner11}, mapping the diffusion potential by means of the local electric field of a scanning tunneling microscope (STM)~\cite{shi-chao12}, and determining the kinetic pathways of adatoms and vacancies by aberration-corrected transmission electron microscopy~\cite{hong17}. 

Scanning probes are routinely used to characterize structural, electronic and mechanical surface properties via imaging, spectroscopy and indentation. Significant progress has been made identifying and moving individual atoms along the surface with atomic force microscopy (AFM) and STM~\cite{custance09,hla14}. However, there is no experimental work mapping the surface diffusion potential by shifting a known single atom along the surface. This is because of the lack of an obvious direct way to extract the diffusion energy barriers from a scanning probe observable. In AFM, dissipative interactions due to internal losses in the force sensor or due to the tip-sample interaction, are accessible through recording the cantilever damping signal~\cite{pishkenari15}. In the absence of structural changes on the tip, contrast in the damping channel can be associated with atomic dynamic processes on the surface which lead to energy dissipation. For example, it has been shown recently that diffusion energy barriers can be estimated from the force sensor's dissipated energy when reshaping C$_{60}$ nanostructures~\cite{freund16}. The relation between tip stimulated diffusion and the energy dissipation has also been discussed in the case of Si adatoms on Si(111)~\cite{arai18}. Using a conducting AFM setup allows to compensate the contact potential, thus preventing electrostatic forces which could trigger undesirable charge related effects.

It is known that on metal and semiconductor surfaces, adatom diffusion can be stimulated by the AFM tip through a lowering of the energy barriers~\cite{kurpick99,custance09} due to the chemical reactivity of the tip with the adatom~\cite{sugimoto08,sugimoto13,sugimoto14,yurtsever17}. Theoretical calculations have demonstrated that the presence of the tip does indeed lower the diffusion barriers and can promote directed diffusion by trapping the adatom, also known as atom gating~\cite{enkhtaivan17}. In the case of the reconstructed Si(001) surface, flipping of the buckled Si dimers driven by a tip induced lowering of the energy barrier is a well-known dissipation channel in AFM imaging~\cite{perez99,kantorovich06,sweetman11,bamidele12}. Naturally, such structural surface modifications are accompanied by atomic scale modifications of the diffusion barriers.

Here, we discuss two schemes to access element specific surface diffusion energy barriers by mapping the energy dissipated in the sensor of an AFM. We start by showing how we can probe the diffusion barriers in constant frequency shift ($\Delta f$) AFM images of self-assembled Mn chains on reconstructed Si(001) by using the Si dimer flipping as a reference. We then focus on the characteristic features in the damping channel during stimulated diffusion of a single Mn atom on reconstructed Si(001). All experiments discussed here were performed at 78 K, thus the condition $kT<E_A$ ($E_A$: activation energy) preventing spontaneous adatom diffusion is valid and we assume $E_A\leq E_{diss}$ ($E_{diss}$: dissipated energy) for tip stimulated single-atom diffusion. We analyze the damping signal by using the well-known Si dimer flip as a reference to compare with other dissipative processes, thus avoiding any assumption regarding the tip other than a stable apex. Using the Si dimer flip as a reference enables the semi-quantitative analysis proposed here. Without such a reference, a very large data set would be required to average out the dependence of the damping signal on the atomistic structure of the tip apex, which is usually not known in details.

\section{\label{sec:level2}Experimental Methods}

Manganese atoms were deposited in-situ on clean p-type (boron doped, 0.1~$\Omega$cm) reconstructed Si(001) surfaces. The deposition was performed in UHV (base pressure $\sim$10$^{-11}$~mbar) at room temperature prior to the STM/AFM investigations. All STM/AFM micrographs reported here were obtained in UHV at 78~K using an Omicron LT-STM. We used qPlus sensors with electrochemically etched W tips and a resonance frequency of $f_0 \approx 25$~kHz. The sensor is equipped with a separate wire for tunneling current to avoid cross-talk. The tip oxide is first removed by scanning the Au(111) surface, followed by controlled tip crashes on Si(001) which very likely lead to a Si tip apex. All STM topographies were acquired in constant $\left<I_t\right>$, while all AFM topographies were taken in constant $\Delta f$, with a constant oscillation amplitude $\approx$ 560 pm and with zero bias. 

\section{\label{sec:level3}Results and Discussion}

\begin{figure}
\begin{center}
\includegraphics[width=\columnwidth]{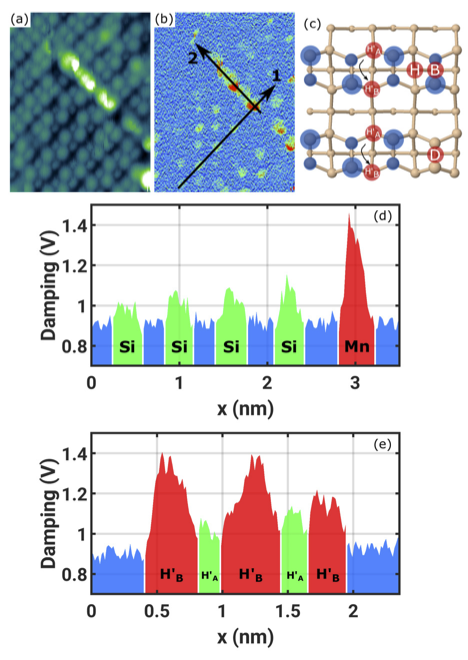}
\caption{(a) AFM topography of a Mn chain on Si(001) at 78 K (3.6 $\times$ 5.0 nm$^2$, $\Delta f$ = -9.7 Hz, slow scan direction is from right to left). (b) Damping channel during the scanning of micrograph (a). (c) Toy model of the Mn chain on Si(001) describing the atom hops seen in (a,b), the Si dimers buckling is represented by the difference in size of the blue circles. Site B is an atom centered over the Si dimer and site D is the replacement of a Si dimer with a single Mn atom. (d) Profile 1 in (b): across the Mn chain. (e) Profile 2 in (b): along the Mn chain. The profiles are color coded as in the damping micrograph in (b).} 
\label{fig:MgngDffsn}
\end{center}
\end{figure}

Fig.~\ref{fig:MgngDffsn}(a) is an AFM topographic image of the Si(001) surface with a single Mn chain running diagonally through the frame. The image was acquired at a constant $\Delta f$ = -9.7 Hz. The Mn atoms in the chain appear blurred, seemingly occupying two sites in the chain structure. This behaviour is unexpected and indicative of atom manipulation~\cite{trevethan07}. Indeed, high resolution STM imaging of stable Mn chains is routinely obtained at 78 K~\cite{villarreal15} and Mn chains are thermodynamically stable below room temperature (break up has been reported in chains heated to 115$^{\circ}$C~\cite{nolph10}). High resolution imaging of the Mn chains is also possible by AFM at an approriate feedback loop setting~\cite{villarreal15}. The motion of each Mn atom composing the chain in Fig.~\ref{fig:MgngDffsn}(a) is driven by the AFM tip as a result of an increased tip-surface interaction. The latter is known to trigger flipping of the Si dimers on the Si(001) surface at 78 K, in particular when sitting over the down Si of the buckled Si dimers~\cite{kantorovich06}. The energy driving this flipping is supplied by the AFM tip and manifests itself as an increased dissipation when the tip is passing over the down Si atom. Similarly, the increased dissipation signal observed in Fig.~\ref{fig:MgngDffsn}(b) when the tip is passing over the Mn chain is a consequence of the AFM tip induced switching of the Mn atom positions between the H'$_A$ and H'$_B$ sites (Fig.~\ref{fig:MgngDffsn}(c)). Hence, the contrast in Fig.~\ref{fig:MgngDffsn}(b) can be considered as a measure of the energy required to move Mn atoms within the chain structure. Note that the movement is determined by the local atomic structure and not the scan direction, suggesting the possibility to extract information about the surface diffusion potential. 

The damping signal can be analyzed more quantitatively by comparison to theoretical calculations of the energy required to induce switching of buckled Si dimers. The latter is estimated by density functional theory (DFT) to be in the range of 100-255~meV for hole doped Si(001)~\cite{ren16}. Fig.~\ref{fig:MgngDffsn}(b) and the profiles in Figs.~\ref{fig:MgngDffsn}(d) and~\ref{fig:MgngDffsn}(e) show that the AFM damping amplitude is similar at the Mn H'$_A$ sites and over the Si-down sites. In agreement with this observation, theoretical calculations find an energy barrier of the order of 160 meV~\cite{wang10, villarreal15} to move a Mn atom from its stable H'$_A$ site in the chain to the H'$_B$ site. This correspondence provides justification to use the AFM damping signal as a measure of the surface diffusion potential. When positioned over the unstable H'$_B$ site of this particular Mn chain, the AFM tip is subject to a significant larger damping. Although better statistics are needed for a robust quantitative analysis, our experiment suggests an increase of the order of 30$\%$ of the dissipation over the H'$_B$ site in comparison to the H'$_A$ site, a number to be compared with future theoretical simulations.

\begin{figure}
\begin{center}
\includegraphics[width=\columnwidth]{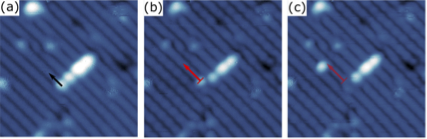}
\caption{STM images (8.6 $\times$ 8.6 nm$^2$) acquired between AFM manipulation sequences of a single Mn atom on Si(001) at 78 K. (a) Initial Mn chain before the Mn atom extraction (black arrow indicates the direction of the AFM path for extraction at constant $\Delta f$). (b) Mn chain and single Mn atom extracted at $\Delta f$ = -28.6~Hz. (c) Mn chain and single Mn atom after stimulated diffusion at $\Delta f$ = -28~Hz. Red arrows indicate the AFM path during the constant $\Delta f$ Mn atom manipulation. STM imaging setpoint was 1.8~V and 20~pA.}
\label{fig:MnpltnTps}
\end{center}
\end{figure}

\begin{figure}
\begin{center}
\includegraphics[width=\columnwidth]{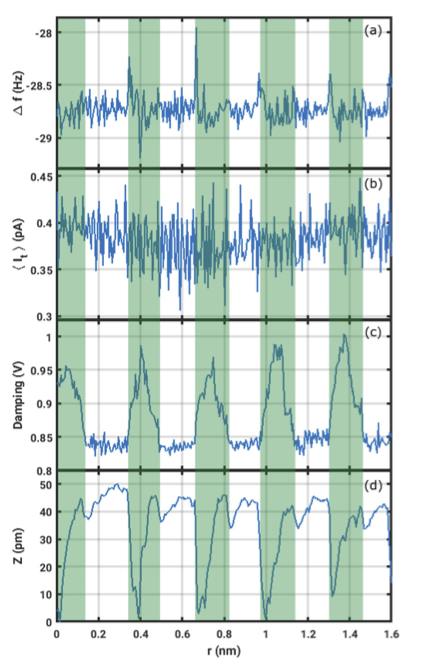}
\caption{Profiles during AFM Mn atom manipulation shown in Fig.~\ref{fig:MnpltnTps}(b) at constant $\Delta f$ = -28 Hz and V$_{LCPD}$ = 110 mV. Sections of non constant damping are highlighted in green. (a) $\Delta f$ channel. (b) $\left<I_t\right>$ channel. (c) Damping channel. (d) Z channel. }
\label{fig:MnpltnPrfls}
\end{center}
\end{figure}

So far, we have demonstrated that the AFM damping signal contributes to understanding the stability of a self-assembled atomic structure. However, we still gain no insight on the chain formation dynamics. To this end, we propose to explore the stimulated diffusion of an isolated Mn atom combining AFM manipulation and STM imaging. The idea is to measure the local AFM dissipation while dragging a single Mn atom along the bare Si(001) surface. To be sure we are indeed moving a Mn atom, we start by extracting one from an existing Mn chain following the sequence illustrated in Fig.~\ref{fig:MnpltnTps}. First, we identify a short Mn chain in a STM micrograph (Fig.~\ref{fig:MnpltnTps}(a)). Next, we proceed with an AFM manipulation in the direction of the black arrow at constant $\Delta$f = -28.6 Hz to isolate one apex Mn atom (Fig.~\ref{fig:MnpltnTps}(b)). This individual Mn atom is then moved further away (Fig.~\ref{fig:MnpltnTps}c) along the red arrow in Fig.~\ref{fig:MnpltnTps}(b) at $\Delta$f = -28 Hz. The latter AFM manipulation path away from any chain provides valuable insight on the diffusion potential of Mn single atoms along the clean Si(001) surface (Fig.~\ref{fig:MnpltnPrfls}). 

During the AFM tip stimulated diffusion, the Mn atom moves along the center of the Si dimer row to a location different from its stable H' edge position in the Mn chain. Three sites are compatible with the STM contrast of the isolated Mn atom in Fig.~\ref{fig:MnpltnTps}(c), where the dark grooves correspond to the center of the Si(001) dimer rows:~\cite{sena11} the H site between two Si dimers, the dimer vacancy filling D site, or the short bridge B site above a Si dimer (see Fig.~\ref{fig:MgngDffsn}(c)). The best agreement with DFT simulated STM images is found for the B site, although energetically it is 500 meV less favorable than the H' site~\cite{wang10}. Further details about the diffusion process can be inferred from the AFM response during the Mn atom motion (Fig.~\ref{fig:MnpltnPrfls}). The atom manipulation is done at constant frequency shift (Fig.~\ref{fig:MnpltnPrfls}(a)) and the tunneling current, monitored simultaneously with a low bias voltage applied to the conducting AFM tip to compensate the local contact potential, is featureless (Fig.~\ref{fig:MnpltnPrfls}(b)). Meanwhile, the AFM dissipation (Fig.~\ref{fig:MnpltnPrfls}(c)) and amplitude (Fig.~\ref{fig:MnpltnPrfls}(d)) reveal strong periodic variations, whose periodicities correspond to the Si dimer spacing ($\approx$3.8 \AA). 

The AFM traces in Fig.~\ref{fig:MnpltnPrfls} provide compelling evidence that the stimulated diffusion process is driven by force interactions between the tip and the Mn atom. The increased dissipation highlighted in green in Fig.~\ref{fig:MnpltnPrfls}(c) is unambiguously linked to the local reduction of the energy barrier to move the Mn atom past each Si dimer, as described in ref~\cite{sugimoto14}. It is indeed only observed when moving a Mn atom along the surface, never while scanning the bare Si(001) surface using the same settings. These dissipation peaks are not compatible with Si dimer flipping, which should only occur every second Si dimer, when the AFM tip is passing over a down Si atom along the red arrow in Fig.~\ref{fig:MnpltnTps}(c).  
Although we cannot image the Mn atom during the tip induced diffusion, the shape and relative phases of the AFM traces in Fig.~\ref{fig:MnpltnPrfls} provide some information on the diffusion process itself. The dissipation signal in Fig.~\ref{fig:MnpltnPrfls}(c) peaks just before the tip drops by about 50 pm towards the surface in Fig.~\ref{fig:MnpltnPrfls}(d). Because we observe the same Z deflection amplitude of the AFM tip when imaging a stable Mn adatom on the Si(001) surface, the tip extension most likely corresponds to the Mn atom jumping away from under the tip. The measured dissipation is similar to the Mn atom on the H' site in the chain and to the Si dimer, providing the basis for future quantitative theoretical modeling.

The multichannel signature in Fig.~\ref{fig:MnpltnPrfls} shows that the Mn atom is being dragged by the AFM tip from H' to B via non-equivalent sites. This is in agreement with theoretical predictions of the lowest energy path for a Mn atom diffusing along the Si dimer row~\cite{wang10,hortamani06}. Attempting to move the Mn atom perpendicular to the Si-dimer rows was very erratic and mostly failed. The above observations suggest Mn atoms are primarily diffusing along the Si dimer rows during the self-assembly process, with a vanishing probability of diffusing perpendicular to them. Such a unidirectional surface diffusion potential landscape can explain the limited length of self-assembled Mn chains, as observed in experiments.  

\section{Conclusion}
In summary, we describe a proof of principle to map the surface diffusion potential of Mn on the clean Si(001) reconstructed surface by measuring the position dependent dissipation of an AFM cantilever during the tip stimulated diffusion of Mn atoms. While theoretical input is required to describe the exact atomistic mechanisms involved in this stimulated diffusion process, some quantitative insight is possible by comparing with the dissipation associated with the well-documented tip induced Si dimer flipping. The experiment shows that Mn is preferably diffusing along the center of the Si(001) dimer rows, with a potential barrier preventing diffusion perpendicular to the dimer rows. This strong anistropy can explain the limited length of self-assembled Mn chains on the Si(001) surface, since additional Mn atoms can only be supplied along a single Si dimer row closest to the Mn chain apex. The generic scheme introduced here can be extended to other adsorbate/surface systems, allowing an original and potentially quantitative insight on adsorption and self-assembly processes.

\begin{acknowledgments}
We thank G. Manfrini and A. Guipet for their technical assistance. C. R., R. V. and A. S. acknowledge support from the Swiss National Science Foundation through division 2 and C. J. K. acknowledges financial support from a UCL Impact Studentship.
\end{acknowledgments}



%

\end{document}